\documentclass[aps,prb,twocolumn,superscriptaddress,showpacs]{revtex4}
\usepackage{epsfig}
\usepackage{amsmath}

\begin{document}

\title{A level coupled to a 1D interacting
reservoir : A DMRG study}
\author{M. Sade, Y. Weiss, M. Goldstein and R. Berkovits}
\affiliation{The Minerva Center, Department of Physics, Bar-Ilan University,
  Ramat-Gan 52900, Israel}

\begin{abstract}

The influence of interactions in a reservoir coupled to
a level on the width of the filling as a function of the chemical potential
and the position of the level is studied.
The density matrix renormalization group (DMRG) method is used
to calculate the ground state of a finite-size interacting reservoir, 
linked to a single state dot.
The influence of the interactions in the lead as well as dot-lead 
interactions  is considered.
It is found that interactions in the reservoir result in a decrease in
the resonance width, while the dot-lead interaction has an opposite
effect. These effects are explained within the random phase approximation as
an effective change in the inverse compressibility of the reservoir,
while the dot-lead interactions renormalize
the position of the level.

\end{abstract}


\maketitle

\begin{figure}[h]\centering
\epsfxsize7cm\epsfbox{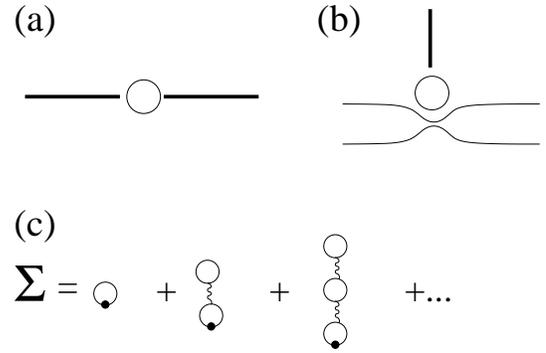}\caption{\label{fig_1}
(a) A level (quantum dot) coupled to two Luttinger liquid 
leads represented by the wide lines. (b)  A level
coupled electrostatically to a QPC
through which the conductance in measured, and to an interacting
reservoir. (c) The diagrammatic representation of the
RPA approximation of the self energy. The line corresponds to
the lead Green function, the black dot to the hopping into the
dot and the wiggly line to the interaction. 
}
\end{figure}

There has been much interest in the conduction through a 1D interacting
system, especially in clarifying 
the behavior of a Luttinger liquid with impurities \cite{kane92}. 
Essentially, it was
shown that any impurity will lead to an insulating behavior. 
The resonance conductance through a
quantum dot coupled to a pair of Luttinger liquid leads 
(see Fig. \ref{fig_1}a) was found to produce
infinitely sharp Coulomb blockade peaks at zero temperature \cite{nazarov03}. 
Thus, no level broadening of the
dot states is exhibited in the measurement of the conduction through
that dot. 

Nevertheless, this does not imply that coupling a dot to a Luttinger liquid 
has no effect on
the width of the filling as a function of the chemical potential.
Consider for example
the arrangement depicted in Fig. \ref{fig_1}b. A dot is connected to
a Luttinger liquid lead, while its occupation 
is measured by a quantum point contact (QPC).
Thus, the Luttinger liquid acts as a reservoir for the dot, while the 
QPC is used to probe the dots level broadening.
In such an arrangement, any additional broadening of the levels due to the
coupling to the reservoir will be seen in the shape of the conductance 
through the QPC.

The difference between the two arrangements is that while the 
first case (Fig. \ref{fig_1}a) essentially probes the 
enhancement of the backscattering in the
Luttinger liquid in the vicinity of the Fermi energy, the second
(Fig. \ref{fig_1}b) explores the broadening of the level due
to coupling to states which may be far from the Fermi energy. 
Therefore, one might expect the broadening of the level 
measured in the second arrangement to approach
the conventional Briet-Wigner form, although some signature of
the interactions is anticipated.

In this paper we study the broadening of a level
coupled to a Luttinger liquid reservoir (Fig. \ref{fig_1}b).
We use the numerical 
density matrix renormalization group (DMRG) \cite{white93} in
order to calculate the ground state of the dot-lead system
for spinless interacting electrons. It will be shown that
the level broadening even for an interacting lead
is well described by the Briet-Wigner form. Despite that,
the interactions in the lead leave a clear signature
in the broadening of the level. If there is no
electrostatic coupling between the dot and lead, the interactions
in the lead result in a reduced width of the level.
On the other hand, such electrostatic coupling leads
to an increase in the width as well as to a change in
the levels' energy.

The DMRG method is used to calculate the orbital population in a system 
consisting of a one-orbital dot
coupled to an interacting lead. At the first stage interactions 
between the dot and the lead
are not taken into account.
The Hamiltonian describing the system is therefore given by:
\begin{eqnarray} \label{eqn:H_no_dl}
{\hat H} &=& \epsilon_{0}{\hat a}^{\dagger}{\hat a} 
-V ({\hat a}^{\dagger}{\hat c}_{1} + 
{\hat c}^{\dagger}_{1}{\hat a}) -t
\displaystyle \sum_{j=1}^{N-1}({\hat c}^{\dagger}_{j}{\hat c}_{j+1} + h.c.) 
\\ \nonumber
&+&I \displaystyle \sum_{j=1}^{N-1}({\hat c}^{\dagger}_{j}{\hat c}_{j}
{\hat c}^{\dagger}_{j+1}{\hat c}_{j+1}),
\end{eqnarray}
where $\epsilon_0$ is the dot's orbital energy level, 
$V$ ($t$) is the dot-lead (lead) hopping matrix element, and
$I$ is the nearest-neighbor interaction strength
in the lead. ${\hat a}^{\dagger}$ (${\hat a}$) is the creation (annihilation)
operator of an electron in the dot, 
and ${\hat c}_j^{\dagger}$ (${\hat c}_j$) is the creation (annihilation)
operator of an electron at site $j$ in the lead.

The Hamiltonian $H$ was diagonalized using a finite-size DMRG calculation 
\cite{white93,berkovits03}
for a lead of 150 sites, and
$n(\mu)$ curves were calculated for several coupling strengths, $V$, 
and interaction strengths, $I$. 
The lead hopping element, $t$, is taken as $1$, 
in order to set the energy scale.

\begin{figure}[h]\centering
\epsfxsize6.5cm\epsfbox{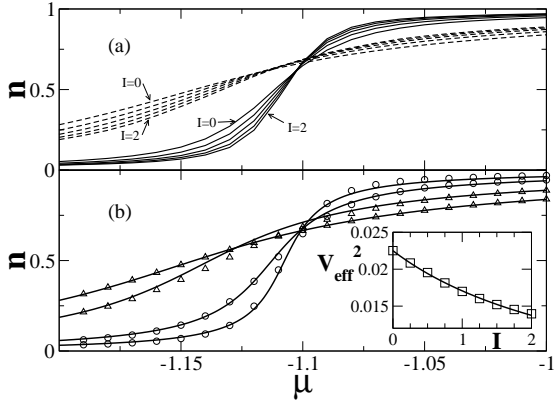}\caption{\label{fig:1a}
(a) The dot's population as a function of the chemical potential for different 
interaction strengths. The full lines represents results for lead-dot coupling 
$V=0.15$, and the dashed lines for $V=0.3$. 
In both cases the interaction strength
$I$ takes values between $0$ and $2$, in jumps of $0.5$. 
(b) The same plots for $I=0$ and $2$ are
drawn again ($V=0.15$ in circles, $V=0.3$ in triangles) 
together with theoretical fits using Eq.(\ref{eqn:moshe1}) (lines).
Inset: $V_{eff}^2$ as a function of $I$ (symbols) as 
obtained by fitting the $n(\mu)$
curves of $V=0.15$ to Eq.(\ref{eqn:moshe1}). 
The line corresponds to the dependence according to 
Eq.(\ref{eqn:interaction1}).}
\end{figure}

Typical results for $n(\mu)$ are shown in Fig.~\ref{fig:1a}. As can be seen,
increasing the interaction strength in the lead ($I$) results in a 
decrease of the level width, while almost no shift in the level position
occurs.

In order to estimate the influence of the interactions in the
lead on the shape of $n(\mu)$, one should start from considering 
the non-interacting case. The coupling of the dot state to the continuum
(akin to the Fano-Anderson model) may be treated using standard
Green function technique \cite{mahan} which leads to:
\begin{eqnarray} \label{eqn:n_mu}
{n}_{dot}(\mu) = \frac{1}{\pi} \int_{-\infty}^{\mu} \frac
{\Im \Sigma(\epsilon)}
{(\epsilon-\epsilon_0-\Re \Sigma(\epsilon))^2 + (\Im \Sigma(\epsilon))^2} 
d\epsilon,
\end{eqnarray}
where $\Sigma(\epsilon)$ is the self energy given by:
\begin{eqnarray} \label{eqn:self_energy}
\Sigma(\epsilon) = \sum_k \frac {|V_k|^2}{\epsilon - \epsilon_k - i \delta},
\end{eqnarray}
$\epsilon_k$ are the eigenvalues of the lead, $V_k$ is the coupling
between the eigenstates in the lead and the state in the dot and 
$\delta \rightarrow 0$.

For the idealized case, the density of states in the lead is
constant (i.e., $\epsilon_k=k / L \nu$, where 
$\nu$ is the (constant) local density of states, and $L$ is the leads length). 
The coupling is $V_k=\sqrt{a/L}V$ ($a$ is the nearest neighbor distance), and 
under these conditions
\begin{eqnarray} \label{eqn:self_energy}
\Sigma(\epsilon) = \frac{a}{L} 
\int \frac {|V|^2 dk}{\epsilon - k/L\nu - i \delta},
\end{eqnarray}
resulting in $\Im \Sigma(\epsilon) = \pi a \nu |V|^2 = \Gamma/2$ and
$\Re \Sigma(\epsilon)=0$. Thus, one obtains
the Breit-Wigner formula: 
\begin{eqnarray} \label{eqn:breit_wigner}
{n}_{dot}(\mu) = \frac{1}{\pi} \int_{-\infty}^{\mu} \frac
{\frac{\Gamma}{2}}
{(\epsilon-\epsilon_0)^2 + (\frac{\Gamma}{2})^2} d\epsilon.
\end{eqnarray}

For the tight-binding model given in Eq.(\ref{eqn:H_no_dl}),
$\epsilon_k= -2t \cos(ka)$ and $V_k=\sqrt{2a/L} \sin(ka) V$, resulting in
$\Im \Sigma(\epsilon) = (V^2/t) \sqrt{1-(\epsilon/2t)^2}$ and
$\Re \Sigma(\epsilon) = (V/t)^2\epsilon/2$. We thus find
\begin{eqnarray} \label{eqn:moshe1}
{n}_{dot}(\mu) = \frac{1}{\pi} \int_{-2}^{\mu} \frac {\frac{V^2}{t} 
\sqrt {1-\frac{\epsilon^2}{4t^2}} }
{\frac{V^4}{t^2} (1-\frac{\epsilon^2}{4t^2}) + 
((1-\frac{V^2}{2t^2})\epsilon - \epsilon_0)^2} d\epsilon.
\end{eqnarray}

We shall now turn to discuss the role played by interactions in the
lead. It is well known that the excitations in the vicinity
of the Fermi energy of any 1D interacting system 
should be described as a Luttinger liquid.
Nevertheless, the dot occupation ${n}_{dot}(\mu)$ is determined by
contributions from all energies, and the region around the Fermi energy
does not play a unique role. Therefore, one could expect a simple perturbation
description of the interactions in the lead to suffice. Indeed, the
effect of the e-e interactions in the RPA approximation
on the self energy (see Fig. \ref{fig_1}c) 
may be written as \cite{berkovits97}:
\begin{eqnarray} \label{eqn:interaction}
\Sigma(\epsilon) = \chi \Sigma^0(\epsilon) 
\end{eqnarray}
where $\Sigma^0(\epsilon)$ is the non-interacting self energy and
\begin{eqnarray} \label{eqn:interaction1}
\chi = \frac{1}{1+a \nu I}.
\end{eqnarray}

Here we
assumed a constant local density of states $\nu$ (for the
tight binding lead $\nu = (a \pi t)^{-1}$ which ignores
local density of states  variations), thus one obtains
$\Im \Sigma(\epsilon) = (V^2/t(1+I/\pi t)) 
\sqrt{1-(\epsilon/2t)^2}$ and
$\Re \Sigma(\epsilon) = (V^2/t(1+I/\pi t))\epsilon/2$, corresponding
to replacing $V^2$ in
Eq.(\ref{eqn:moshe1}) by an ``effective'' coupling 
$V_{eff}^2=V^2(1+I/\pi t)$.

Returning to the results obtained by the numerical DMRG calculations,
the curves of Fig.~\ref{fig:1a} can be now fitted to 
Eq.(\ref{eqn:moshe1}) with two fitting parameters -
$\epsilon_{0,eff}$ and $V_{eff}^{2}$.
It is easy to see that the Briet-Wigner form
fits quite well
even in the presence of strong interactions.
The effect of interactions is limited
here to a decrease of $V_{eff}^{2}$, i.e., decrease of $\Gamma$,
while the level position $\epsilon_{0,eff}=\epsilon_0$ 
remains constant.
The values of $V_{eff}^{2}$ extracted from the fit are plotted
in Fig.~\ref{fig:1a}(inset) and compared with the RPA predictions
of $V_{eff}^2=V^2(1+I/\pi t)$. A rather good correspondence is
observed.

In order to consider interactions between an electron occupying 
the dot, and the electrons in the lead, an additional interaction
term should  be added to the Hamiltonian
(Eq. (\ref{eqn:H_no_dl})):
\begin{eqnarray} \label{eqn:H_dl}
{\hat H_{dl}} = I_{dl} {\hat a}^{\dagger}{\hat a}
{\hat c}^{\dagger}_{1}{\hat c}_{1}.
\end{eqnarray}

\begin{figure}[h]\centering
\epsfxsize6.5cm\epsfbox{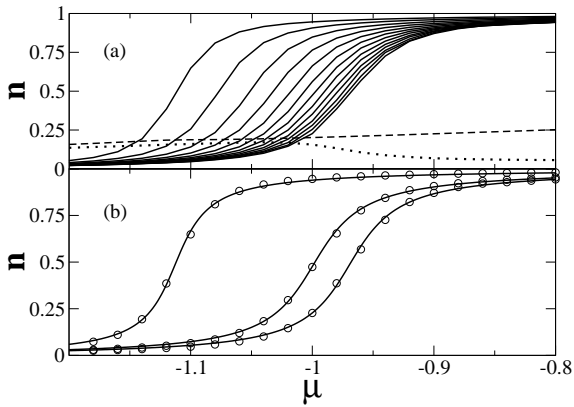}\caption{\label{fig:2a}
(a) Population of the dot and of the first site of the lead 
as a function of the chemical potential, for $V=0.15$. 
Dot-lead interaction was included (Eq.(\ref{eqn:H_dl})) taking $I_{dl}=I$.
The curves shown are for $I$ between $0$ and $3$, in jumps of $0.25$ (full lines, dot population),
and for $I=0$ (dashed line) and $I=2$ (dotted line) for the lead population.
(b) The plots for $I=0, 1.5$ and $3$ (symbols)
together with the best fit to Eq.(\ref{eqn:moshe1}). }
\end{figure}

DMRG calculations were performed for $H+H_{dl}$, and the corresponding 
$n(\mu)$ results (Fig.~\ref{fig:2a})
clearly show a change in the resonance width, but also a
change in the level position, which was absent in
the previous case. Nevertheless, these results can still be
fitted to Eq.(\ref{eqn:moshe1}) with the same fitting 
parameters - $\epsilon_{0,eff}$  and $V_{eff}^{2}$.
As can be seen, Eq.(\ref{eqn:moshe1}) describes this system quite well.

\begin{figure}[h]\centering
\epsfxsize6.5cm\epsfbox{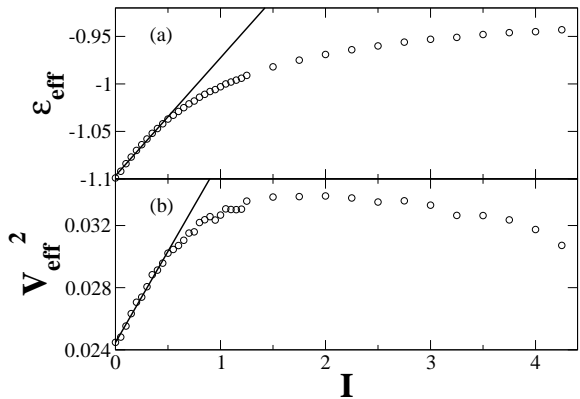}\caption{\label{fig:2b}
(a) $\epsilon_{0,eff}$ and (b) $V_{eff}^{2}$ as functions of 
$I_{dl}=I$ for $V=0.15$ 
(symbols) and linear fits for the region $I\le0.5$ (lines). }
\end{figure}

The resonance center movement, $\epsilon_{0,eff}$, as well as the
width, $V_{eff}^{2}$, that were obtained from the fit
can be seen in Fig. \ref{fig:2b}.
For small values of interaction
both grow linearly.  First lets try to explain the shift in the
resonance center. As noted, almost no shift was seen for $I_{dl}=0$.
In the presence of weak dot-lead interactions, one may approximate
${\hat H_{dl}} \sim I_{dl} n_1 {\hat a}^{\dagger}{\hat a}$, 
where $n_{1}$ represents the average occupation of the first
site in the lead. As can be seen in Fig.~\ref{fig:2a}a, $n_1$ is not
very sensitive to the occupation of the dot and may be replaced
by its typical value. Thus, the energy of the orbital in 
Eq.(\ref{eqn:H_no_dl}) may be rewritten as
$\epsilon_{0,eff} = \epsilon_{0}+ n_1 I_{dl}$. Indeed, 
this formula fits well the numerical results for small 
values of $I_{dl}$ (as can be seen
in Fig.~\ref{fig:2b}(a)), for $n_{1} = 0.14$. This result agrees well
with the value $n_{1} \sim 0.15$ in the region of the resonance,
taken from the data of Fig.~\ref{fig:2a}(a).

A more striking feature is the behavior of $V_{eff}^{2}$, i.e,
the width of the resonance. There is a distinct qualitative 
change in the width behavior,
compared to the case without dot-lead interactions.
As opposed to the monotonic decrease of $V_{eff}^{2}$, which 
was demonstrated in Fig.~\ref{fig:1a} (inset), 
Fig.~\ref{fig:2b}b (symbols) shows that $V_{eff}^{2}$ increases with $I$, 
until a maximal value is achieved around $I=2$. For larger values of
interaction decrease in the width is observed.

This enhancement of $V_{eff}^{2}$ is associated to the interplay
between the population of the dot level to the depopulation
of the first site in the lead, ignored in our treatment of $\epsilon_{0,eff}$.
This leads to a reduction in the effect of the dot-lead interaction
which results in an increase in the width as depicted 
in Fig.~\ref{fig:2b}b. For weak interactions the enhancement 
of  $V_{eff}^{2}$ is linear.

Thus, although the Luttinger liquid
has a vanishing local DOS at the end of the lead
in the vicinity of the Fermi energy,
a level coupled to a 1D interacting reservoir is 
broadened, 
since all the reservoir states take part in the broadening mechanism.
Nevertheless, as we have seen, the interactions in the reservoir influence the
width of the resonance. One might gain some insight
from the following consideration:
For the non-interacting case
(for constant density of states in the reservoir)
the width is equal to $\Gamma = 2 \pi a \nu |V|^2$, which
may be rewritten  as $\Gamma = 2 \pi (a/L) |V|^2 \partial N/ \partial \mu$.
The thermodynamic inverse compressibility 
$\partial N/ \partial \mu$ is affected by the interactions \cite{lee82}.
Lets consider the compressibility $\partial \mu/ \partial N$.
In the lowest order approximation \cite{berkovits97}
$\partial \mu/ \partial N = (L \nu)^{-1} + e^2/C$, where $C$ is the capacitance
of the system. For nearest neighbor interaction $e^2=aI$ and
as usual $C \sim L$. Therefore,  
$\partial N/ \partial \mu = L \nu /(1+a \nu I)$. Inserting this to
the expression for $\Gamma$, we get a result similar to
the RPA approximation results in Eq.(\ref{eqn:interaction1}).
Although capacitance is proportional to the length $L$ of the lead,
so is the density of states in a 1D system, and therefore
it has an influence even for an infinite lead.

The conductance through the QPC in the geometry
described in Fig. \ref{fig_1}b
is directly proportional to the occupation of the dots' orbital due to
the capacitive coupling between the charge of the dot and 
the QPC (it is assumed that no tunneling occurs between the dot and
the QPC) \cite{johnson04}. Thus, in principal, ${n}_{dot}(\mu)$, may be
read off the conductance through the QPC and the effect of coupling 
of the dot to the interacting reservoir can be measured.

In conclusion, interactions in a reservoir coupled to
a resonant level leave clear fingerprints on the
width and position of the resonance.
The main influence is the decrease in
the resonance width due to a
change in the inverse compressibility of the reservoir.
On the other hand, the dot-lead interaction shifts the resonance position
and may also enhance the width.

Support from the Israel Academy of Science (Grant 276/01) is gratefully
acknowledged.

\end{document}